\documentclass[conference]{IEEEtran}
\ifCLASSOPTIONcompsoc
  \usepackage[nocompress]{cite}
\else
  \usepackage{cite}
\fi

\ifCLASSINFOpdf
\else
\fi

\author{\IEEEauthorblockN{Hisham Alasmary$^\diamond$, Aminollah Khormali$^\diamond$, Afsah Anwar$^\diamond$, Jeman Park$^\diamond$, \\Jinchun Choi$^{\diamond\ddagger}$, DaeHun Nyang$^\ddagger$ and Aziz Mohaisen$^\diamond$} \\
\vspace{-2mm}
\IEEEauthorblockA{$^\diamond$University of Central Florida \hspace{3mm} $^\ddagger$Inha University}
\hspace{3mm}  \\ \{hisham, aminkhormali, afsahanwar, parkjeman, jc.choi\}@knights.ucf.edu \\ nyang@inha.ac.kr, amohaisen@gmail.com}

\usepackage{color, amsmath, balance, xcolor, float, lscape, enumerate, graphicx, url, tabularx, multirow, xspace, hyperref, keyval, adjustbox,subcaption,wrapfig,booktabs,tikz,amssymb}

\usepackage{wrapfig}

\newcommand{\BfPara}[1]{{\noindent\bf#1.}\xspace}
\usepackage{tikz}

\usepackage[final]{pdfpages}
\usepackage{hyperref}
\definecolor{linkcolour}{rgb}{0,0.2,0.6}
\definecolor{xgreen}{rgb}{0.2,0.6,0.0}
\definecolor{xred}{rgb}{0.7,0.1,0.0}

\tikzset{near start abs/.style={xshift=0.75cm, yshift=0.75cm}}
\usetikzlibrary{matrix,chains,positioning,decorations.pathreplacing,arrows}

\begin{document}

\title{Analyzing, Comparing, and Detecting Emerging Malware: A Graph-based Approach}

\maketitle

\begin{abstract}
The growth in the number of Android and Internet of Things (IoT) devices has witnessed a parallel increase in the number of malicious software (malware), calling for new analysis approaches. We represent binaries using their graph properties of the Control Flow Graph (CFG) structure and conduct an in-depth analysis of malicious graphs extracted from the Android and IoT malware to understand their differences. Using 2,874 and 2,891 malware binaries corresponding to IoT and Android samples, we analyze both general characteristics and graph algorithmic properties. Using the CFG as an abstract structure, we then emphasize various interesting findings, such as the prevalence of unreachable code in Android malware, noted by the multiple components in their CFGs, and larger number of nodes in the Android malware, compared to the IoT malware, highlighting a higher order of complexity. We implement a Machine Learning based classifiers to detect IoT malware from benign ones, and achieved an accuracy of 97.9\% using Random Forests (RF).
\end{abstract}

\vspace{-3mm}

\section{Introduction}\label{sec:introduction}\vspace{-2mm}

\BfPara{Background} As IoT finds new applications, IoT software security becomes of a paramount importance. IoT malware stands as one of the most significant threats to the security and stability of the Internet, and understanding IoT malware through analysis and detection is an essential problem to mitigate their security threats~\cite{Gerber17,Harrison15}. The limited existing literature on IoT malware, and despite malware analysis, classification, and detection being a focal point of analysts and researchers~\cite{MohaisenAM15,ShangZXXZ10,AlasmaryA0CNM18}, points at the difficulty, compared to other malware types. To understand IoT malware, we perform software analysis between IoT and Android samples, using graph properties obtained from CFG structures, and build a detection system of IoT malware utilizing those properties.

\BfPara{Overview} Starting with a new dataset of IoT malware samples, we pursue a graph-theoretic approach to malware analysis. Each malware sample can be abstracted into a Control Flow Graph (CFG) to extract representative static features of the application. As such, graph-related features from the CFG can be used as a representation of the software, and classification techniques can be built to tell whether the software is malicious or benign. Using the CFG graph constructs, We perform a comparative study of those graph-theoretic features in both types of software to highlight the CFG shift in IoT malware to Android application malware to uncover various similarities and differences. We similarly analyze the CFGs of 261 IoT benign samples and use that to build IoT classifiers from 23 different features extracted from the CFGs.

\vspace{-1mm}

\section{Dataset} \vspace{-2mm}
Our IoT malware set is 2,874 samples, randomly selected from IoTPOT~\cite{PaSYM2016}. We also obtained a dataset of 2,891 Android malware samples from~\cite{ShenVMKZ17} for contrast. Finally, we manually created a dataset of benign samples from source files gathered from {\em OpenWrt.org}~\cite{openwrt}, and kernel files. To this end, we disassembled the IoT binaries, in the form of Executable and Linkable Format (ELF) files, as well as the Android Application Packages (APKs) using {\em Radare2}~\cite{radare2} to extract the CFG from the disassembly codes. Moreover, we used an off-the-shelf tool, NetworkX~\cite{networkx2013}, for further graph analysis. 
\vspace{-1mm}

\section{Evaluation Metrics and Results}\label{sec:method}\vspace{-1mm}
\BfPara{Evaluation Metrics} For our initial analysis of the various malware (Android and IoT) and benign samples, we use various standard algorithmic graph properties, including the number of nodes, the number of edges, the closeness, the number of components, etc. For the lack of space, we omit the definitions of those properties, and refer the interested reader to~\cite{AlasmaryA0CNM18} for more details. In the following, we use a normalized version, from 0 to 1, of the closeness centrality.
\vspace{-2mm}

\subsection{Analysis} \vspace{-1mm}

\begin{figure}[t]
\begin{minipage}{0.24\textwidth}
     \centering        \includegraphics[width=0.99\textwidth, height=1.25in]{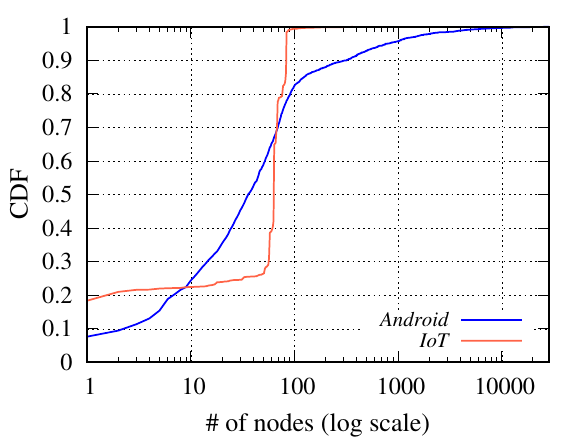}
        \caption{Log scale for nodes}\vspace{-4mm}
        \label{fig:node}
\end{minipage}
\begin{minipage}{0.24\textwidth}
        \centering
        \includegraphics[width=0.99\textwidth, height=1.25in]{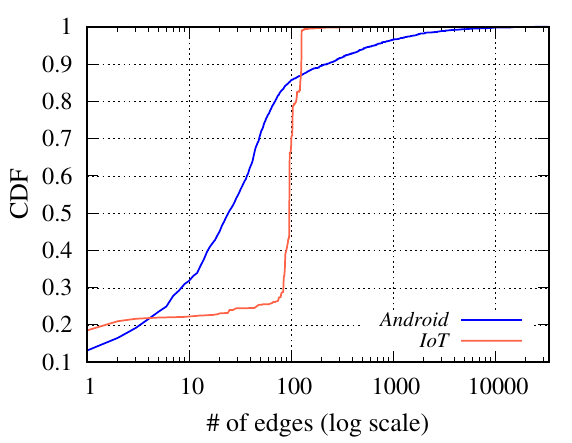}
        \caption{Log scale for edges}\vspace{-4mm}
        \label{fig:edge}
\end{minipage}
\end{figure}

\BfPara{Android malware size differs from IoT malware significantly} Upon analyzing CFG of different samples belonging to each class, we observed that the Android and IoT malware samples have at least 28,691 and 367 nodes, and 33,887 and 577 edges, respectively. Figure~\ref{fig:node} and Figure~\ref{fig:edge} represent the logarithmic scale for the number of nodes and edges, where the dynamic region of the CDF in Figure~\ref{fig:node} is between 1 and 60 nodes, while the active region in Figure~\ref{fig:edge} between around 1 to 85 edges correspond to around [0.2--0.3] (about 10\% of samples).  This combined finding of the number of edges and nodes in itself is very intriguing: while the number of nodes in IoT malware samples is relatively smaller than that in Android malware, the number of edges is higher. This is striking, as it highlights a simplicity at the code base (smaller number of functions) yet a higher complexity at the flow-level (more edges; calls between functions), adding a unique analysis angle to the malware that is only visible through CFG structure.

\begin{figure}[t]
\begin{minipage}{0.24\textwidth}
    \centering
    \includegraphics[width=0.99\textwidth]{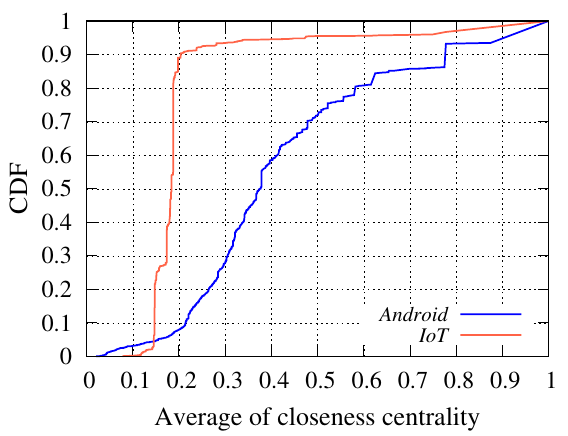}
    \caption{Closeness centrality in the largest components.}\vspace{-4mm}
    \label{fig:closeness}
\end{minipage}
\begin{minipage}{0.24\textwidth}
    \centering
    \includegraphics[width=0.99\textwidth,height=1.25in]{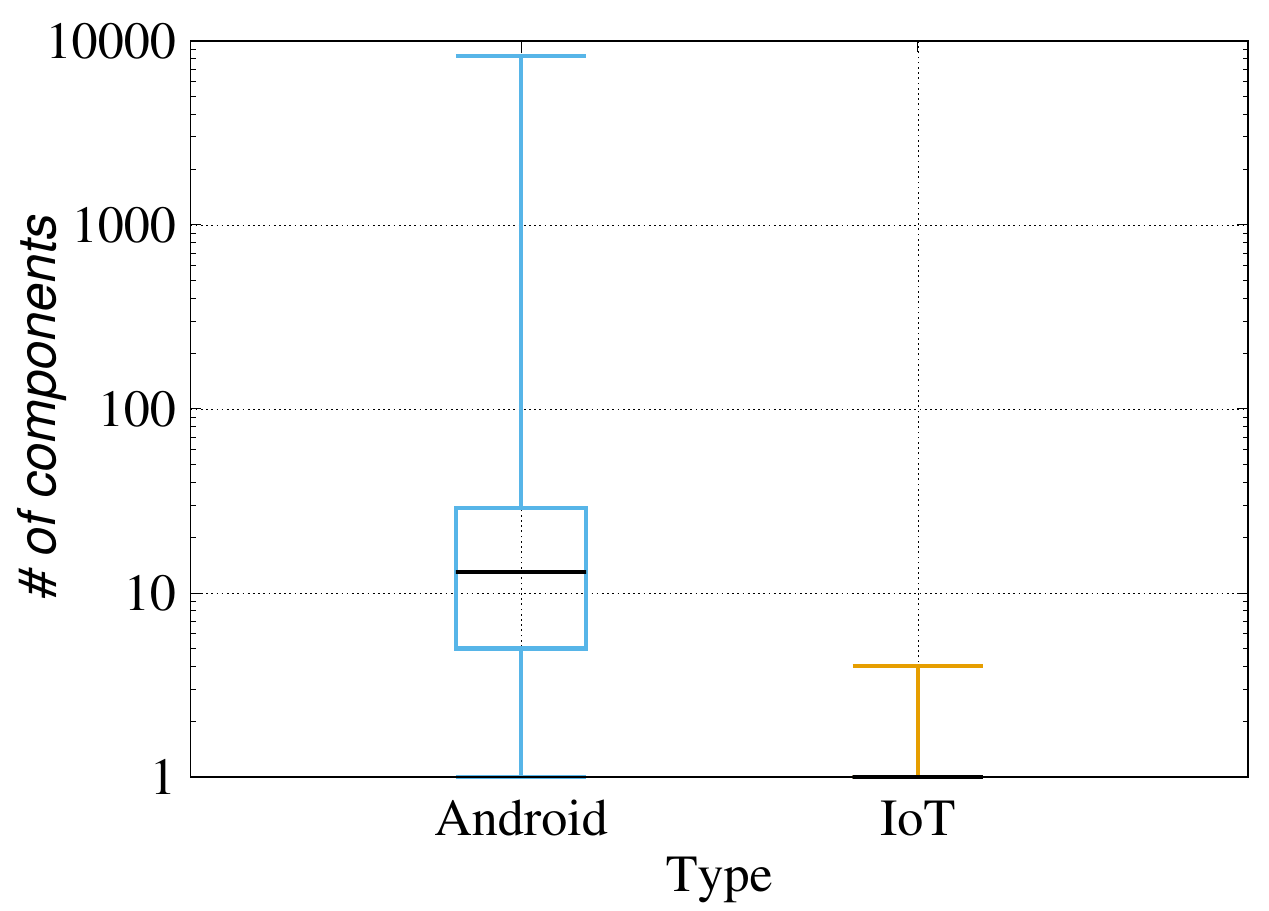}
    \caption{Number of components: IoT vs. Android.}\vspace{-5mm}
    \label{fig:components}
\end{minipage}\vspace{-2mm}
\end{figure}

\BfPara{IoT CFG's are not only dense, but also well enmeshed graphs} Figure~\ref{fig:closeness} and Figure~\ref{fig:components} depict the CDF for the average closeness centrality and number of components, respectively, for both datasets. To reach this plot, we notice that around 5\% of the IoT and Android have around 0.14 average closeness centrality. On the other hand, The same 80\% of IoT samples have a closeness of less than 0.19, highlighting that the closeness alone with the value 0.2 can be used as a distinguishing feature of the two different types of the malware. The relatively higher value also highlight that IoT graphs are well enmeshed.

\BfPara{CFG analysis shed light on software anomalies} 3.23\% of the IoT malware (93 IoT samples) have more than two components; i.e., most have one component that have file sizes from 56,500 -- 266,200 bytes. On the other hand, 13.83\%, or 400 Android samples, have only one component, where their size ranges from around 4,200 -- 9,400,000 bytes. However, 2,491 samples (around 86.17\%) have more than one component, which show that the Android malware often uses unreachable functions. We observe multiple components in Android CFG, which show the presence of multiple entry-points in the same program. These point the use of decoy functions with the aim to circumvent an analyst when trying to analyze the malware. The gap between datasets can be noticed, showing the new shift trend of the Android malware to the IoT devices.

\vspace{-2mm}
\subsection{Classification} \vspace{-2mm}
As a result of the differences between IoT and Android malware across those graph features, it's natural to utilize those features of classification. To this end, we build a classifier for detecting IoT malware against bengin IoT samples. 
Upon extracting 23 different features from the CFGs for all samples (based on the betweenness centrality, closeness centrality, degree centrality, shortest path, density, \# of edges, and \# of nodes). Upon initial analysis, we obtained 2,347 IoT samples for classification against 261 benign samples. The results are reported in~\autoref{tab:class_malicious} using standard binary classification performance metrics. The results are obtained using 10-fold cross-validation. As shown, we obtained an accuracy of 97.87\% using Random Forest classifier.
\vspace{-2mm}

\begin{table}[t]
\centering
\caption{Classification results for the whole dataset, biased towards malicious samples. All results are percentages. False Negative Rate (FNR), False Positive Rate (FPR), False Discovery Rate (FDR), False Omission Rate (FOR), F1 score (F1), and Accuracy Rate (AR). } \vspace{-1mm}
\scalebox{0.9}{
\begin{tabular}{|l|l|l|l|l|l|l|l|l|}
\hline

Method               & \multicolumn{2}{c|}{Actual} & \multicolumn{1}{c|}{FNR} & \multicolumn{1}{c|}{FPR} & \multicolumn{1}{c|}{FDR} & \multicolumn{1}{c|}{FOR} & \multicolumn{1}{c|}{F1} & \multicolumn{1}{c|}{AR} \\ \hline
\multirow{2}{*}{LR}   & 16.6        & 6.7         & \multirow{2}{*}{28.5} & \multirow{2}{*}{4.0} & \multirow{2}{*}{36.3} & \multirow{2}{*}{2.9} & \multirow{2}{*}{67.0} & \multirow{2}{*}{93.8} \\ \cline{2-3}
                     & 9.5         & 228.0       &                        &                       &                        &                       &                    &    \\ \hline
\multirow{2}{*}{SVM}   & 22.3        & 6.1         & \multirow{2}{*}{20.7} & \multirow{2}{*}{1.6} & \multirow{2}{*}{14.5} & \multirow{2}{*}{2.6} & \multirow{2}{*}{81.8} & \multirow{2}{*}{96.2} \\ \cline{2-3}
                     & 3.8         & 228.6       &                        &                       &                        &                       &                    &   \\ \hline
\multirow{2}{*}{RF}  & 23.6        & 3.1         & \multirow{2}{*}{11.6} & \multirow{2}{*}{1.1} & \multirow{2}{*}{9.6}  & \multirow{2}{*}{1.3} & \multirow{2}{*}{89.5} & \multirow{2}{*}{97.9} \\ \cline{2-3}
                    & 2.5         & 231.6       &                        &                       &                        &                       &                      &   \\ \hline
\multirow{2}{*}{CNN}  & 22.9        & 3.0        & \multirow{2}{*}{1.3} & \multirow{2}{*}{11.5} & \multirow{2}{*}{1.4}  & \multirow{2}{*}{1.4} & \multirow{2}{*}{98.7} & \multirow{2}{*}{97.6} \\ \cline{2-3}
                     & 3.2         & 231.7       &                        &                       &                        &                       &                    &    \\ \hline
\end{tabular}}\vspace{-4mm}
\label{tab:class_malicious}
\end{table}

\section{Conclusion and Future Work}\label{sec:conclusion} \vspace{-1mm}
We conduct an in-depth graph-based analysis of three different datasets to highlight the similarity and differences of IoT and Android malware, as well as benign IoT software towards detection of new IoT malware. Toward this goal, we extract the CFGs as an abstract representation to characterize IoT malware across different graph features, and highlight the shift in the graph representation from the IoT to the Android malware by tracing size (nodes, edges, and components). We observe decoy functions for circumvention, which correspond to multiple components in the CFG. Using those features, we built a classifier that achieved 97.9\% of accuracy with 1.1\% FPR and 11.6\% FNR in IoT malware detection.

\BfPara{Acknowledgement}
This work was supported by NSF CNS-1809000, NRF-2016K1A1A2912757, Florida Cybersecurity Center (FC2) seed grant, and NVIDIA GPU Grant Program.
\vspace{-2mm}

\balance
\bibliographystyle{IEEEtran}
\bibliography{ref,conf}

\end{document}